# Off-axis electron holography of magnetic nanostructures: magnetic behavior of Mn rich nanoprecipitates in (Mn,Ga)As system

M. Barańska[1], P. Dłużewski[1], S. Kret[1], K. Morawiec[1], Tian Li[1], J. Sadowski[1,2,3]

[1] Institute of Physics Polish Academy of Sciences, al. Lotników 32/46, 02-668 Warsaw, Poland

[2] MAX-IV laboratory, Lund University, P.O. Box 118, 221 00 Lund, Sweden

[3] Department of Physics and Electrical Engineering, Linnaeus University, 391 82 Kalmar, Sweden

e-mail: mbaranska@ifpan.edu.pl

## Abstract

The Lorentz off-axis electron holography technique is applied to study the magnetic nature of Mn rich nanoprecipitates in (Mn,Ga)As system. The effectiveness of this technique is demonstrated in detection of the magnetic field even for small nanocrystals having an average size down to 20 nm.

75.50.Pp Magnetic semiconductors

**Introduction**

The main difficulties encountered when using TEM to study magnetic materials is that the sample is usually immersed in a large magnetic field of the objective lens. It is enough to completely eliminate or severely distort most of the domain structures of interest. Many strategies have been developed to overcome the high field problem in the sample region. In our case it is turning off the standard objective lens, to provide a field-free region for the sample and we use the off-axis electron holography, which is one of the modes of magnetic TEM (Lorentz microscopy), to obtain information about magnetic structure. In general, studies can be made of specimens in their as-grown state, in remanent states, in the presence of applied fields or currents, and as a function of temperature.

The Lorentz electron holography is a very powerful technique for mapping static electric and magnetic fields in nanoscale. There are many interesting experimental and theoretical variations of this technique [1]. For example, the off-axis electron holography using an electrostatic biprism [2, 3], the in-line electron holography involving the reconstruction of through-focus series [4, 5] or the hybrid electron holography which is a combined method of the two aforementioned ones [6, 7]. All of these approaches have their advantages and limitations.

The off-axis electron holography in the Lorentz mode is the most suitable method for investigation of an internal magnetic field in nanoscale. One of the most commonly used approach needs at least two holograms made for a specimen in the upside and downside configurations. The acquired holograms are the subject of a multistep computer processing, which requires consideration of several experimental conditions crucial for the final separation of the phase shifts caused by the electrostatic and magnetic fields.

Many nanostructures, such as nanowires [8, 9, 10], nanocubes [11, 12], nanospheres [13, 14], nanoparticles [15, 16] or nanorings [17] have been investigated by means of electron holography in view of their magnetic nature. In this paper we present the off-axis electron holography approach to the study of the magnetic behavior of Mn rich nanoprecipitates embedded in (Mn,Ga)As matrix [18, 19]. The sample is tested both at room temperature and at low temperatures. Both the composition and the size of the nanoparticles represent the novelty compared to others studies.

## Off - axis electron holography

The off-axis electron holography may be used to recover the amplitude and the phase of an electron wave function that passed through a TEM sample. The phase of the electron wave function is sensitive to both the electrostatic and the magnetic fields of the sample.

The electrostatic contribution must be removed from the reconstructed phase images in order to reach the magnetic contribution to the phase shift of an object. One of several ways to achieve this separation is to record two holograms for the same region of the specimen in normal (upside) and reversal (downside) positions. The sum and the difference of the reconstructed phase images can then be used to supply the electrostatic and the magnetic contribution to the phase, respectively [20].

Three coupled experimental parameters affect the quality of the final reconstructed phase image significantly: overlap width, interference fringe spacing and interference fringe contrast. In general, the application of higher biprism voltage results in a larger overlap width, a smaller interference fringe spacing and a weaker fringe contrast.

**Results and discussion**

The off-axis electron holograms of Mn rich precipitates embedded in (Mn,Ga)As matrix were collected with utilization of FEI Titan 80-300 TEM equipped with biprism wire. The biprism voltage of 250 V and 10 s exposure time were used for acquiring the holograms. All images were recorded on a 2048 x 2048 pixel CCD camera. The off-axis holograms were reconstructed by means of Holo3 [21] and HolograFree [22] software. The reconstructed phases were exploited to extract the electrostatic and magnetic contributions to the phase shift.

In this paper, four holograms for the same region of the sample are analyzed: one for the upside and one for the downside configuration at room temperature (RT) of 296 K and analogous holograms for liquid nitrogen temperature (LNT) of 120 K.

Fig.1 presents an example of experimental off–axis hologram showing the interference fringe changes caused by the nanoprecipitates. The resulted phase and amplitude images were obtained by means of Holo3 program [21].

Fig. 2 shows the electrostatic and the magnetic phases determined from the upside/downside method. The top row presents the measurements at RT, while the bottom row - at LNT. As one can notice, the electrostatic phase images for RT and LNT, are similar to each other, while there are clear differences between the magnetic phase images for RT and LNT.

It has also been tested whether the difference between the total phase images obtained for LNT and for RT in the upside or downside configuration would give the same magnetic contribution to the phase shift as the upside/downside method. In this case, it is presumed that the electrostatic phase shift does not vary with temperature.

Fig. 3 shows that the obtained phase images for both configurations look similarly and are comparable with those from the upside/downside method. Based on these results it can be assumed that internal magnetic fields of the nanoprecipitates are increasing in LNT.

During the experiment and analysis of the images a few problems were encountered. In general, the recording of the off–axis electron holograms at high resolution is a very demanding task. This is mostly because of the requirements of long exposure times, high stability of the equipment and sensitivity to vibrations of the experimental setup environment. Moreover, the off-axis electron holography suffers from the noise levels in both the amplitude and the phase and the off-axis reconstruction contains Fresnel fringes and artifacts from the biprism wire surface.

Another problems appear in the case of small nanoparticles (below 10 nm in diameter). It is difficult to find a suitable region of the specimen, where the nanoparticles are clearly visible in both the upside and the downside positions. The reversal of the sample may cause that the nanoparticles are seen in a different way, which can result in significant misalignment of the images leading to inaccurate products of subtraction and addition of phase images. The appropriate alignment of the phase images plays a crucial role in the case of such small nanoparticles, especially for the magnetic phase shift which, in comparison to the electrostatic phase shift, is much weaker and therefore more sensitive to the image misalignment.

**Conclusions**

The utility of the off-axis electron holography in the study of small (down to 20 nm) magnetic nanoobjects was demonstrated. Despite the numerous experimental problems, the manifestation of ferromagnetism for Mn rich nanoprecipitates in (Mn,

Ga) As system was found in the liquid nitrogen temperature.

**Acknowledgments:**

This research was co-financed by the Polish National Science Center (Grant No. UMO-2013/11/B/ST3/04244 and Grant No. 2014/13/B/ST3/04489) and by the European Regional Development Fund within the Innovative Economy Operational Programme 2007-2013 (Grant No. POIG.02.01-00- 14-032/08). We are also grateful for additional support from the EAgLE international project (FP7-REGPOT-2013-1, Project No. 316014) co-financed by Polish Ministry of Science and Higher Education, Grant Agreement 2819/7.PR/2013/2.

**List of figures**

[1] Fig.1. Experimentally acquired off-axis electron hologram of MnAs nanoprecipitates embedded in (Ga,Mn)As matrix (a), the enlarged region of the hologram showing two MnAs precipitates (b), the amplitude of Fourier transform of the electron hologram (c), the reconstructed amplitude image (d) and the unwrapped phase image (e).

[2] Fig. 2. The images of phase shift due to the electrostatic field (left column – a, c) and the magnetic field (right column – b, d) at RT (top row – a, b) and at LNT (bottom row – c, d), obtained with the application of the upside/downside method. All phase

images are 20x enhanced and refer to the isolated area of Fig.1b.

[3] Fig. 3. The results of subtraction of the total phase images obtained at RT and at LNT for upside (a) and for downside (b) specimen positions. Both phase images are 40x enhanced and refer to the isolated area of Fig.1b. The phase shift difference between RT and LNT images is due to the increasing of magnetization of MnAs precipitations with a decreasing temperature.

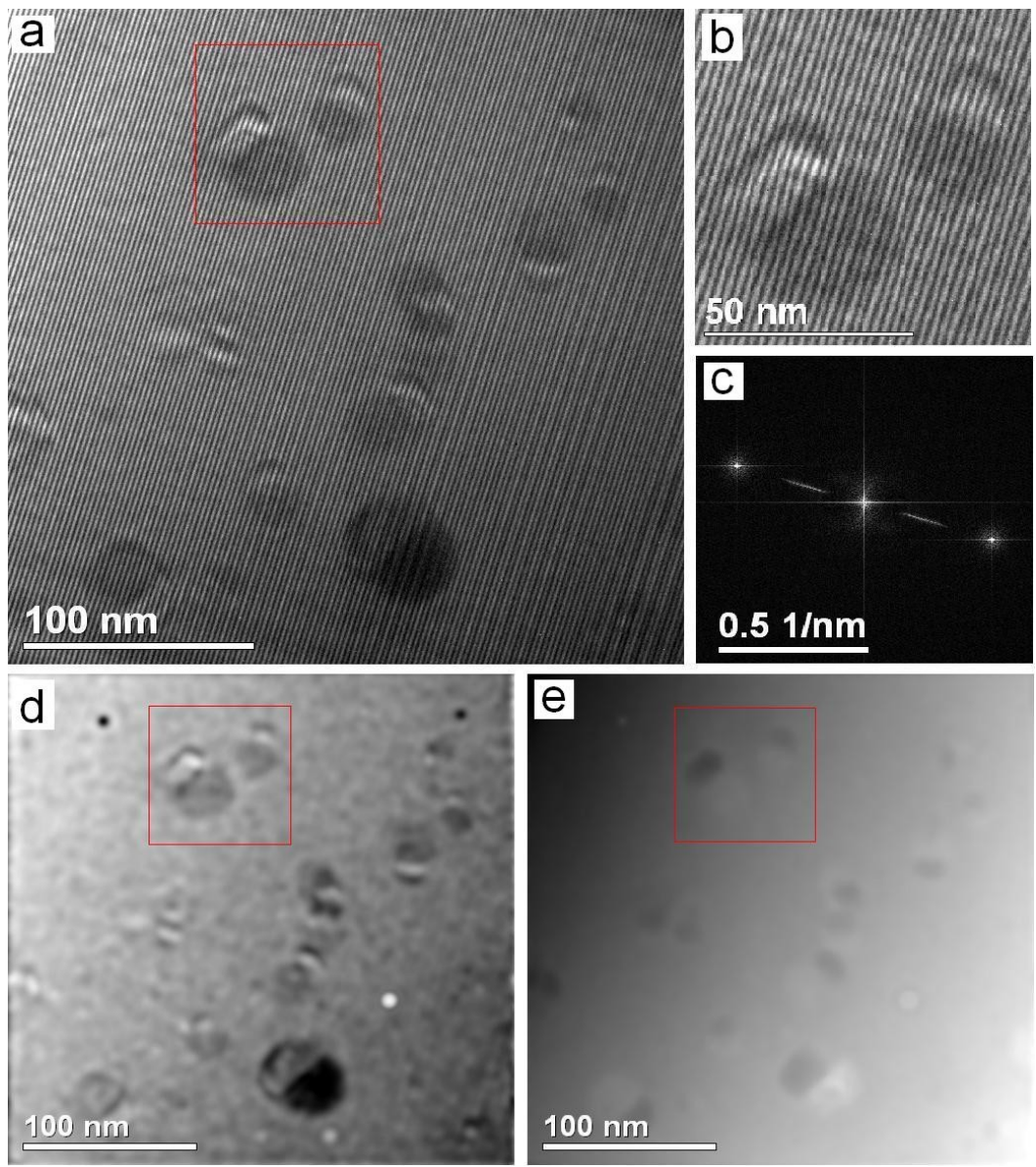

Fig.1. Experimentally acquired off-axis electron hologram of MnAs nanoprecipitates embedded in (Ga,Mn)As matrix (a), the enlarged region of the hologram showing two

MnAs precipitates (b), the amplitude of Fourier transform of the electron hologram (c), the reconstructed amplitude image (d) and the unwrapped phase image (e).

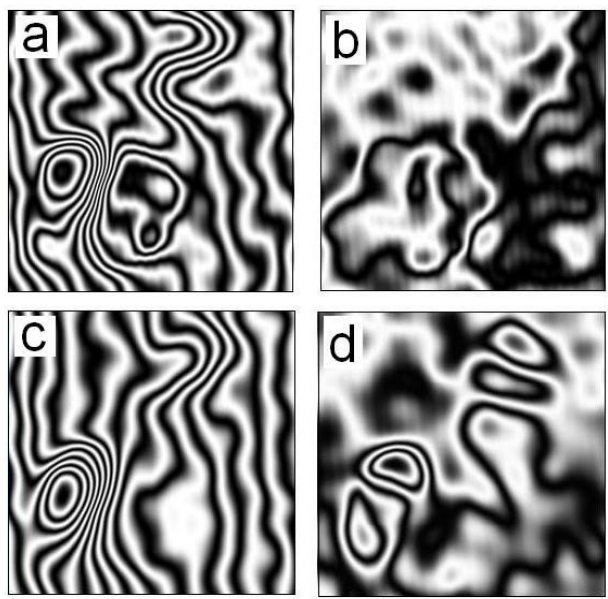

Fig. 2. The images of phase shift due to the electrostatic field (left column – a, c) and the magnetic field (right column – b, d) at RT (top row – a, b) and at LNT (bottom row – c, d), obtained with the application of the upside/downside method. All phase images are 20x enhanced and refer to the isolated area of Fig.1b.

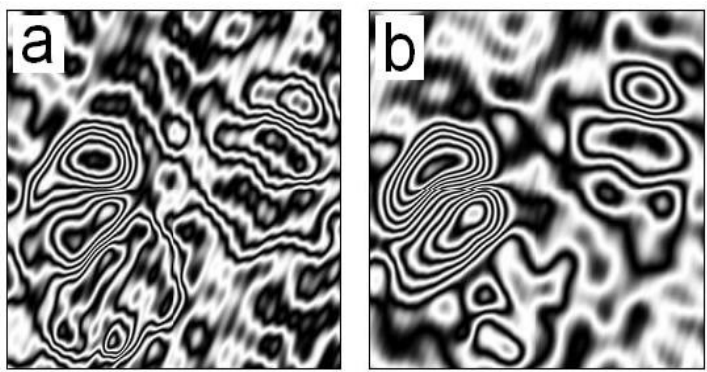

Fig. 3. The results of subtraction of the total phase images obtained at RT and at LNT for upside (a) and for downside (b) specimen positions. Both phase images are 40x enhanced and refer to the isolated area of Fig.1b. The phase shift difference between RT and LNT images is due to the increasing of magnetization of MnAs precipitations with a decreasing temperature.